\documentclass{aa}

\usepackage[varg]{txfonts}

\usepackage{graphicx}	
\usepackage{amsmath}	
\usepackage{subfigure}
\usepackage{epstopdf}
\usepackage[colorlinks, linkcolor=blue, urlcolor=blue, citecolor=blue]{hyperref}

\usepackage{booktabs}
\usepackage{ulem}

\usepackage{threeparttable}
\usepackage{enumitem}
\usepackage{longtable}

\def\be{\begin{equation}}
\def\ee{\end{equation}}

\usepackage{CJK}
\usepackage{upgreek}

\begin{document}

\title{Repeating fast radio bursts from synchrotron maser radiation in localized plasma blobs: Application to FRB 20121102A}

\begin{CJK*}{UTF8}{gbsn}
\author{Xiao Li \inst{\ref{inst1}}
\and Fen Lyu \inst{\ref{inst2},\ref{inst1}}
\and Hai Ming Zhang \inst{\ref{inst1}}
\and Can-Min Deng \inst{\ref{inst1}}
\and En-Wei Liang\inst{\ref{inst1}}}

\institute{Guangxi Key Laboratory for Relativistic Astrophysics, School of Physical Science and Technology, Guangxi University, Nanning 530004, China\label{inst1}
\\
\email{lyufen@aqnu.edu.cn,lew@gxu.edu.cn}
\and
Institute of Astronomy and Astrophysics, School of Mathematics and Physics, Anqing Normal University, Anqing 246133, People's Republic of China
 \label{inst2}}

\date{Received XXX / Accepted XXX}

\abstract{
The radiation physics of repeating fast radio bursts (FRBs) remains enigmatic. Motivated by the observed narrow-banded emission spectrum and ambiguous fringe pattern of the spectral peak frequency ($\nu_{\rm pk}$) distribution of some repeating FRBs, such as FRB 20121102A, we propose that the bursts from repeating FRBs arise from synchrotron maser radiation in localized blobs within weakly magnetized plasma that relativistically moves toward observers. Assuming the plasma moves toward the observers with a bulk Lorentz factor of $\Gamma=100$ and the electron distribution in an individual blob is monoenergetic ($\gamma_{\rm e}\sim300$), our analysis shows that bright and narrow-banded radio bursts with peak flux density $\sim$ 1 ${\rm Jy}$ at peak frequency ($\nu_{\rm pk}$) $\sim 3.85$ GHz can be produced by the synchrotron maser emission if the plasma blob has a magnetization factor of $\sigma\sim10^{-5}$ and a frequency of $\nu_{\rm P}\sim 4.5$ MHz. The spectrum of bursts with lower $\nu_{\rm pk}$ tends to be narrower. 
Applying our model to the bursts of FRB 20121102A, the distributions of both the observed $\nu_{\rm pk}$ and isotropic energy $E_{\rm iso}$ detected by the Arecibo telescope at the L band and the Green Bank Telescope at the C band are successfully reproduced. We find that the $\nu_{\rm P}$ distribution exhibits several peaks, similar to those observed in the $\nu_{\rm pk}$ distribution of FRB 20121102A. This implies that the synchrotron maser emission in FRB 20121102A is triggered in different plasma blobs with varying $\nu_{\rm P}$, likely due to the inhomogeneity of relativistic electron number density.}


\titlerunning{Repeating FRBs from plasma blobs}
\authorrunning{Xiao Li, et al.}

\maketitle

\section{Introduction}\label{sec:intro}
\end{CJK*}
Fast radio bursts (FRBs) are bright radio transients that last several to tens of milliseconds and are mostly extragalactic, with a typical dispersion measure (DM) of $\sim$$100-3038$$\, {\rm {pc}}\,{\rm {cm^{-3}}}$ \citep{2007Sci...318..777L,2012MNRAS.425L..71K,2013Sci...341...53T,2019ARA&A..57..417C,2019A&ARv..27....4P,2022A&ARv..30....2P,2021ApJ...910L..18B,2021ApJS..257...59C}. More than 800 FRBs have been detected to date \citep{2016PASA...33...45P,2021ApJS..257...59C}, over 60 of which exhibit repetitive behaviors \citep{2020ApJ...891L...6F,2023ApJ...947...83C}\footnote{\url{https://blinkverse.zero2x.org/}}. Similar to the spectra of one-off FRBs, the spectra of bursts from individual repeating FRB sources display significant diversity \citep{2016Natur.531..202S,2019ApJ...872L..19M}. However, repeating FRBs typically exhibit longer durations and narrower bandwidths than one-off FRBs \citep{2021ApJS..257...59C,2021ApJ...923....1P}. Bursts from repeating FRB sources often exhibit complex time-frequency drift structures, and some bursts consist of several sub-bursts \citep{2019ApJ...876L..23H,2022RAA....22l4001Z}. The question of whether all FRBs are repeating remains unresolved 
\citep{2019MNRAS.484.5500C,2022ApJ...926..206Z}.
\par 

The origin of FRBs is still unclear and widely debated (see \citealt{2019PhR...821....1P,2023RvMP...95c5005Z} for reviews). Most proposed source models involve some compact objects, such as magnetized neutron stars \citep{2016ApJ...829...27D,2016ApJ...822L...7W},
young pulsars \citep{2016MNRAS.462..941L,2021FrPhy..1624503L}, 
magnetars
\citep{2014MNRAS.442L...9L,2017ApJ...843L..26B,2020ApJ...896..142B,2019MNRAS.485.4091M,2020MNRAS.498.1397L}, strange stars \citep{2018ApJ...858...88Z,2021Innov...200152G}, and black holes \citep{2020MNRAS.494L..64K,2021PhRvD.103l3030D}. The detected association of magnetar SGR 1935+2154 with FRB 20200428 suggests that at least some FRBs may originate from magnetars \citep{2020Natur.587...59B,2020Natur.587...54C}. 

The observed spectral profiles of bursts from repeating FRBs are typically modeled with a Gaussian function \citep{2017ApJ...850...76L,2021ApJ...922..115A,2022RAA....22l4001Z}. These bursts are generally narrow-banded. For instance, the spectra of bursts emitted by FRB 20201124A have a characteristic bandwidth ($\Delta\nu$) of $\sim 0.277$ GHz, with a peak frequency ($\nu_{\rm pk}$) of 1.09 GHz \citep{2022RAA....22l4001Z}. The relative spectral bandwidth $\Delta \nu /{\nu _{\rm pk}} $ for 1076 bursts from FRB 20220912A, detected by the Five-hundred-meter Aperture Spherical Radio Telescope (FAST, \citealt{2019SCPMA..6259502J}), is narrowly distributed in the range $\left( {0.1 \sim 0.2} \right)$ \citep{2023ApJ...955..142Z}. For one burst from repeating FRB 20190711A, $\Delta \nu /{\nu _{\rm pk}}$ is 0.065/1.4 \citep{2021MNRAS.500.2525K}. These observations suggest that the narrow spectra of these typical FRBs are likely intrinsic and determined by their intrinsic radiation mechanism \citep{2023ApJ...956...67Y,2024A&A...685A..87W}.
\par

Multiple frequency observations, particularly wideband frequency observations, are critical for understanding the radiation mechanism of FRBs. FRB 20121102A, the first detected repeating FRB source \citep{2016Natur.531..202S}, has had thousands of bursts reported by various monitoring campaigns across $0.5-8$ GHz (e.g., \citealt{2014ApJ...790..101S,2018ApJ...863....2G,2018ApJ...866..149Z,2019A&A...623A..42H,2019ApJ...882L..18J,2020A&A...635A..61O,2020MNRAS.495.3551R,2021Natur.598..267L,2022MNRAS.515.3577H}). It has a typical $\Delta \nu$ less than 500 MHz \citep{2018ApJ...863....2G,2022ApJ...941..127L}. Moreover, the peak frequency of bursts observed by the Green Bank Telescope (GBT) at the C band ($4-8$ GHz) shows several discrete peaks \citep{2018ApJ...863....2G,2018ApJ...866..149Z,2022ApJ...941..127L}. Interestingly, when extending such a fringe spectral feature to $0.5-4$ GHz, the bimodal burst energy distribution observed with FAST by \cite{2021Natur.598..267L} can be well reproduced by assuming a simple power law energy function \citep{2022ApJ...941..127L}. Similar discrete peaks are also seen in the peak frequency distributions of repeating FRB sources FRB 20190520B and FRB 20201124A \citep{2023MNRAS.522.5600L,2024ApJ...966..115L}.
\par

The high brightness temperature ($\rm {T_B}\geq{10^{35}}\, \rm K$) indicates that the radiation mechanism of FRBs must be coherent \citep{2020Natur.587...45Z,2021Univ....7...56L,2021SCPMA..6449501X}. Various hypotheses have been proposed, such as synchrotron maser radiation in relativistic shocks under strong magnetization conditions \citep{2014MNRAS.442L...9L,2017ApJ...843L..26B,2020ApJ...896..142B,2019MNRAS.485.4091M} or weak magnetization conditions \citep{2017ApJ...842...34W} as well as vacuum conditions
\citep{2017MNRAS.465L..30G}, coherent curvature radiation, coherent inverse Compton scattering, or coherent Cherenkov radiation close in the magnetosphere \citep{2018ApJ...868...31Y,2022ApJ...925...53Z,2023ApJ...958...35L}. Synchrotron maser radiation under weak magnetization conditions has a very narrow intrinsic radiation spectrum and a prominent peak \citep{2002ApJ...574..861S}. Inspired by the observed narrow-banded emission spectrum and the fringe pattern of the $\nu_{\rm pk}$ distribution of repeating FRBs, we explore whether the synchrotron maser radiation mechanism of electrons in a weakly magnetized relativistic plasma can account for these spectral characteristics.
\par

The paper is organized as follows. The model is presented in Sect. \ref{sec:Mod}. The application of our model to explain the spectral characteristics of FRB 20121102A and to constrain the model parameters via Monte Carlo simulations is shown in Sect. \ref{sec: simulation}. Conclusions and discussion are given in Sect. \ref{sec:DIS and SUM}. Throughout the paper, we adopt a flat $\Lambda$CDM cosmology with cosmological parameters $H_{0}$=67.7$\mathrm{~km}$ $\mathrm{~s}^{-1}$ $\mathrm{Mpc}^{-1}$, $\Omega_{m}=0.31$ \citep{2016A&A...594A..13P}.

\section{Model} \label{sec:Mod}
We propose that repeating FRBs arise from a pre-accelerated plasma that relativistically moves toward observers with a bulk Lorentz factor of $\Gamma$. It may originate from a relativistic outflow powered by the central engine (e.g.,  \citealt{2014MNRAS.442L...9L,2017ApJ...842...34W,2019MNRAS.485.4091M,2020ApJ...896..142B,2022ApJ...927....2K}).
 A burst episode results from plasma instability induced by ejecta injected from the activity of the central engine, such as magnetar flares. The ejecta is highly relativistic and may be dominated by Poynting flux or baryons. The interaction between the ejecta and the plasma shell generates collisionless shocks, which induce plasma instability or turbulence and form localized blobs. 
In case of the synchrotron maser emission conditions are satisfied in some blobs, bright FRB events with a narrow-banded spectrum can be generated by the synchrotron maser radiation of electrons in the plasma blobs (e.g., \citealt{2002ApJ...574..861S,2017ApJ...842...34W,2019ApJ...875..126G}).
The observed frequency-dependent depolarization may be due to the FRB emission propagation through the clumpy shell   \citep{2022Natur.609..685X,2022Sci...375.1266F}.
\par

\begin{figure}[h]
	\includegraphics[width=1\columnwidth]{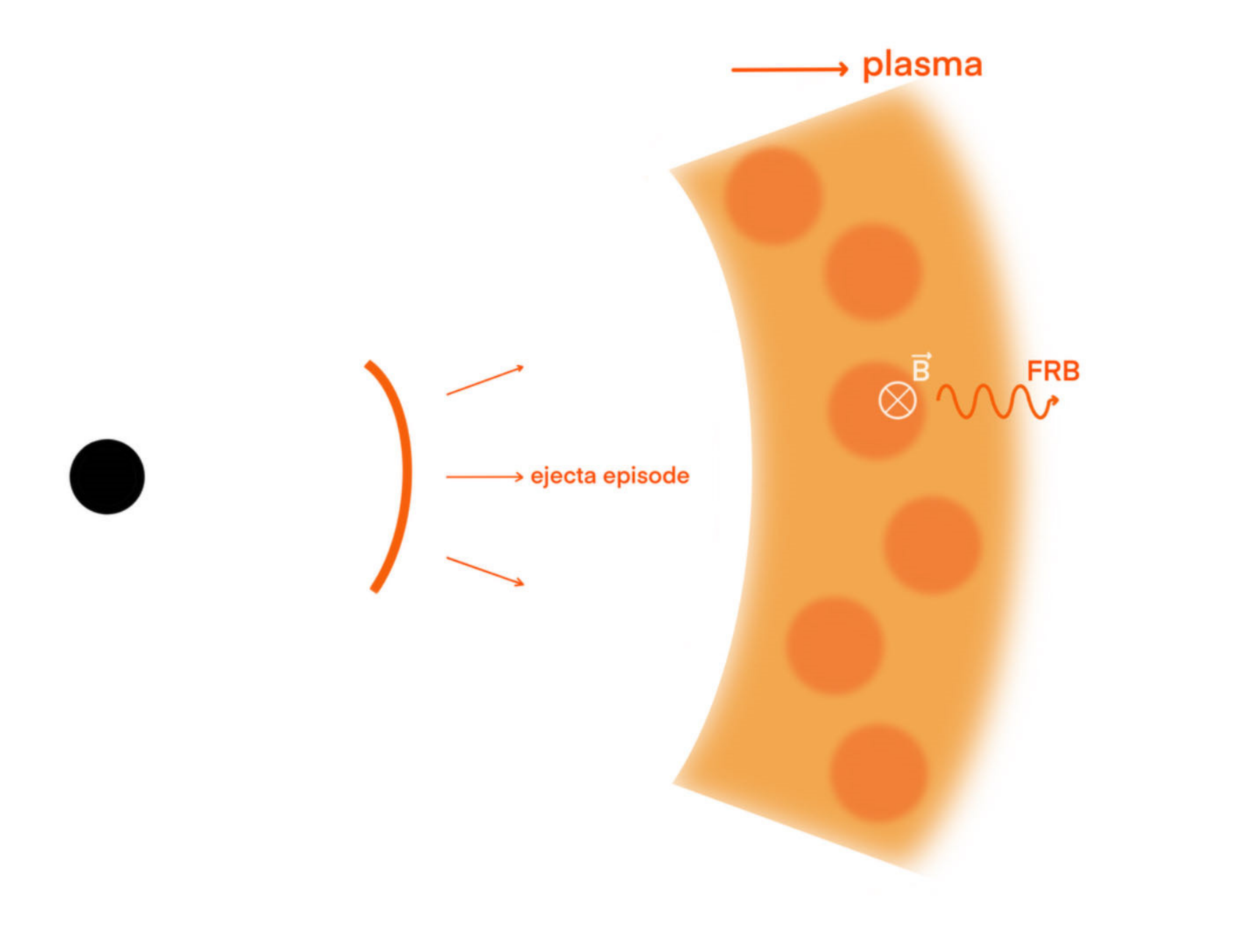}
    \centering
\caption{Schematic configuration: the ejecta from the central engine triggers plasma instabilities, inducing localized electron plasma blobs that generate FRBs.}\label{MyFig_cartoon}
\end{figure}

We show the cartoon of our model in Fig. \ref{MyFig_cartoon}. The size of the blob in the comoving frame is estimated as $R^{'}=\Gamma c \Delta t\sim 3\times 10^{9}$ cm, assuming $\Gamma=100$ and $\Delta t=1$ ms. The plasma frequency is given by ${\nu_{\rm P}} = {\left( {{{n_{\rm e}{{\rm e}^2}}}/{{\pi \gamma_{\rm e} {m_{\rm e}}}}} \right)^{1/2}}$ and the magnetization factor is defined as $\sigma={{ {B^2}}}/{{4\pi {m_{\rm e}}{c^2}\gamma_{\rm e} n_{\rm e}}} = {\left( {{{{\nu _B}}}/{{{\nu_{\rm P}}}}} \right)^2}$, where $n_{\rm e}$ is the relativistic number density of electrons, $\rm e$ is the electron charge, $m_{\rm e}$ is the rest mass of the electron, $B$ is the magnetic field, $\gamma_{\rm e}$ is the Lorentz factor of the electron and $\nu_{ B}$ is the cyclotron frequency of the relativistic electron \citep{2021Univ....7...56L}.
\par

\begin{figure}
	\includegraphics[width=\columnwidth]{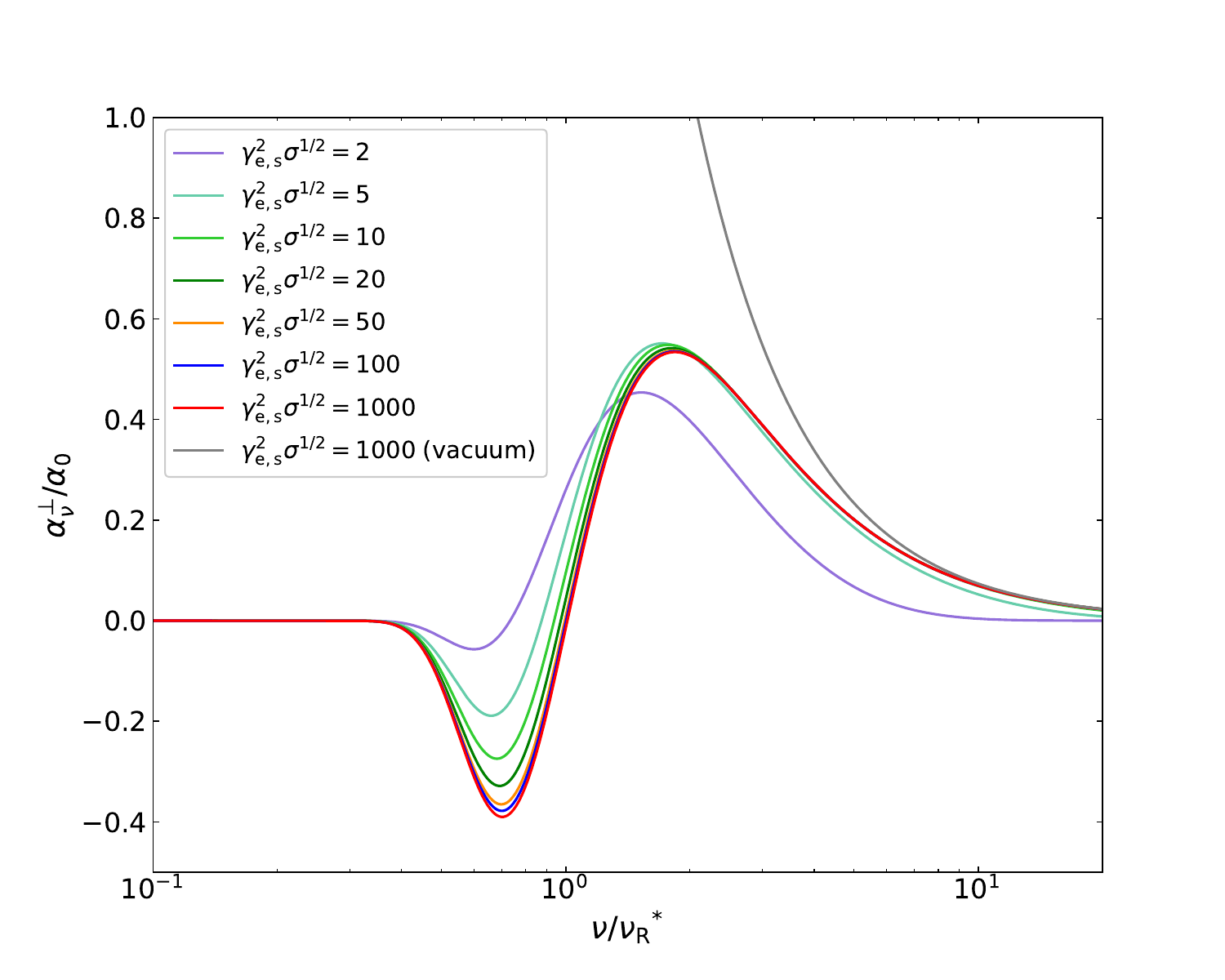}
    \centering
    \caption{Normalized self-absorption coefficient for $ \bot $ polarization, given a single-energy electron distribution and $\chi=\frac{\pi }{4}$, is plotted against frequency normalized by the scaling factor $\nu _{\rm {R^*}}$. The gray solid line represents the results obtained without plasma effects.}\label{MyFig1}
\end{figure}

The synchrotron radiation power of a relativistic electron in the plasma is severely suppressed for emission at $\nu \lesssim \nu _{\rm {R^*}}$. This suppression is known as the Razin effect\footnote{The synchrotron emission beaming angle of a relativistic electron in the plasma is given by $\theta = \sqrt {1 - {\rm n^2}{\beta ^2}}$, where $\rm n$ represents the refractive index (${\rm n^2} \approx 1 - {{\nu_{\rm P}^2}}/{{{\nu ^2}}} \ne 1$) and $\beta$ is the velocity of the electron in units of speed of the light \citep{1986rpa..book.....R}. In the case of $\beta \approx 1$, we have $\theta\approx\sqrt {1 - {\rm n^2}} = {\nu_{\rm P}}/\nu $, indicating that the beaming angle depends on both the emission frequency and plasma frequency. For a certain emission frequency below $\nu_{\rm R}$ (the so-called Razin frequency $\gamma_{\rm e}\nu_{\rm P}$), its beaming angle is significantly larger than that in the vacuum, $ {\nu_{\rm P}}/\nu \gg 1/\gamma_{\rm e}$, resulting in a significant weakening of the beaming effect \citep{1966Sci...154.1320M,1986rpa..book.....R}. Furthermore, \citet{2002ApJ...574..861S} generalized the typical frequency at which the beaming effect is weakened to $\nu\lesssim \nu_{\rm {R^*}}={\nu_{\rm P}}\min \left\{ {\sigma^{-1/4},{\gamma _{\rm e}}} \right\}$, where $\nu _{\rm {R^*}}$ is the modified Razin frequency.}. Another striking effect of synchrotron radiation in the plasma is the maser emission mechanism, which involves the amplification of radiation caused by the negative synchrotron self-absorption \citep{1958AuJPh..11..564T,1966Sci...154.1320M,1967JETP...24..381Z,1970SvA....13..797S,2002ApJ...574..861S,2017ApJ...842...34W}. The self-absorption coefficient for a specific polarization mode of synchrotron radiation is given by 
\citep{1989aetp.book.....G,2002ApJ...574..861S}
\begin{equation}\label{sec2 equ1}
 \alpha _\nu ^{\left[ \perp, \| \right]} = - \frac{1}{{4\pi {m_{\text{e}}}{\nu ^2}}}\int d {\gamma _{\text{e}}}\gamma _{\text{e}}^2P_\nu ^{\left[ \perp, \| \right]}\left( {{\gamma _{\text{e}}}} \right)\frac{d}{{d{\gamma _{\text{e}}}}}\left( {\gamma _{\text{e}}^{ - 2}\frac{{d{n_{\text{e}}}}}{{d{\gamma _{\text{e}}}}}} \right) \;
\end{equation}
and 
\begin{equation}\label{sec2 equ2}
\begin{aligned}
P_\nu ^{\left[ \perp, \| \right]}\left( {{\gamma _{\text{e}}}} \right)=&\frac{\sqrt{3} {\rm {e^3}} { B}\sin \chi}{2 m_{\rm e} c^{2}}\left[1+\left(\frac{\nu_{\rm R}}{\nu}\right)^{2}\right]^{-1 / 2} \frac{\nu}{\tilde{\nu}_{\rm c}} \times\\ &\left[\int_{\nu / \tilde{\nu}_{\rm c}}^{\infty} K_{5 / 3}(z) d z \pm K_{2 / 3}\left(\frac{\nu}{\tilde{\nu}_{\rm c}}\right)\right] \;,
\end{aligned}
\end{equation} where $P_\nu ^{\left[ \perp, \| \right]}\left( {{\gamma _{\text{e}}}} \right)$ is the radiation power per unit frequency emitted by a single electron with Lorentz factor $\gamma_{\rm e}$ in a polarization mode ${\left[ \perp, \| \right]}$. Here, $\bot$ and $\parallel$ denote the linear polarization perpendicular and parallel to the projection of the magnetic field on the plane of observation, respectively. $\chi$ is the pitch angle, $K_{5/3}$ and $K_{2/3}$ are the modified Bessel functions, and \begin{equation}\label{sec2 equ3}{\tilde \nu }_{\rm{c}} = \frac{{3{\rm{e}}B\sin \chi }}{{4\pi {m_{\rm{e}}}c}}\gamma _{\rm{e}}^2{\left[ {1 + {{\left( {\frac{{{\nu _{\rm{R}}}}}{\nu }} \right)}^2}} \right]^{ - 3/2}} \;.
\end{equation}
The contribution of the negative reabsorption comes from the regions where the electron distribution function is steeper than $\gamma _{\rm e}^2$ \citep{2002ApJ...574..861S,2017ApJ...842...34W,2019ApJ...875..126G}. \citet{1966Sci...154.1320M} and \citet{1967JETP...24..381Z} discussed the possibility of negative reabsorption in cold plasma (non-relativistic plasma). However, even without cold plasma, due to the correction of $\rm n$ by relativistic electron plasma, negative reabsorption can occur at $\nu\lesssim \nu_{\rm R}$ when $ \gamma _{\rm e}^2 \ll \sigma^{-1/2}$ \citep{1970SvA....13..797S}. \citet{2002ApJ...574..861S} derived that negative reabsorption can also occur below the frequency $\sigma^{-1/4}{\nu_{\rm P}} $ even when $\gamma _{\rm e}^2 > {\sigma^{-1/2}} > 1$. Finally, it is generally believed that negative reabsorption occurs at $\nu\lesssim \nu_{\rm {R^*}}={\nu_{\rm P}}\min \left\{ {\sigma^{-1/4},{\gamma _{\rm e}}} \right\}$. Following \citet{2002ApJ...574..861S,2017ApJ...842...34W}, negative reabsorption can arise from a narrow electron distribution. In our analysis, we assume that the electron distribution in the blobs is monoenergetic as $\frac{{d{n_{\rm{e}}}}}{{d{\gamma_{\rm e} }}} = {n_{\rm e}}\delta \left( {{\gamma_{\rm e}  } - {\gamma _{\rm {e,s}}}} \right)$. Moreover, FRB 20121102A shows almost $100\%$ linear polarization degree \citep{2018Natur.553..182M}. Therefore, the reabsorption coefficient of the radiation produced by a monoenergetic electron distribution is given as 
\begin{equation}\label{sec3 eq1}
\alpha _\nu ^{\left[  \bot  \right]} = {\alpha _0}{F^{[ \bot ]}}\left[ {\gamma _{{\rm{e,s}} }^2{\sigma ^{1/2}},\frac{\nu }{{\nu _{\rm{R}}^*}}} \right]\;,\quad {\alpha _0} = \frac{{\pi {\nu_{\rm P}}}}{{2\sqrt 3 c}}{\sigma ^{3/4}}\sin \chi  \;.
\end{equation}
The details of the function ${F^{[ \bot ]}}$ are given in Eq. \eqref{appendix 5} in the Appendix. We present the numerical results of our model by considering $\gamma _{{\rm{e,s}}  }^2{\sigma ^{1/2}}$ values of 2, 5, 10, 20, 50, 100, and 1000, with $\chi=\pi/4$. Figure \ref{MyFig1} shows the normalized self-absorption coefficient $(\alpha _\nu ^{\bot}/{\alpha _0})$ as a function of the frequency in units of $\nu _{\rm {R^*}}$ in the comoving frame. It is observed that negative reabsorption begins at $\nu/\nu_{\rm R^*}\sim 0.4$, independent of the $\gamma _{{\rm{e,s}}  }^2{\sigma ^{1/2}}$ values. The peak negative reabsorption and the corresponding frequency are parameter dependent at $\gamma _{{\rm{e,s}}  }^2{\sigma ^{1/2}}<50$, but they approach an asymptotic value of $(\alpha _\nu ^{\bot}/{\alpha _0})=-0.4$ at $\nu  = 0.70\nu _{\rm R^*}$ by increasing the values of $\gamma _{{\rm{e,s}}  }^2{\sigma ^{1/2}}$ up to 1000. The negative reabsorption regime is confined in the frequency range of $0.4\nu _{\rm R^*} < \nu  < \nu _{\rm R^*}$. The gray solid line in Fig. \ref{MyFig1} represents the positive self-absorption coefficient against frequency neglecting the plasma effects (The detailed calculation is shown in Eq. \eqref{appendix 8} in Appendix).

\begin{figure}[h]
	\includegraphics[width=\columnwidth]{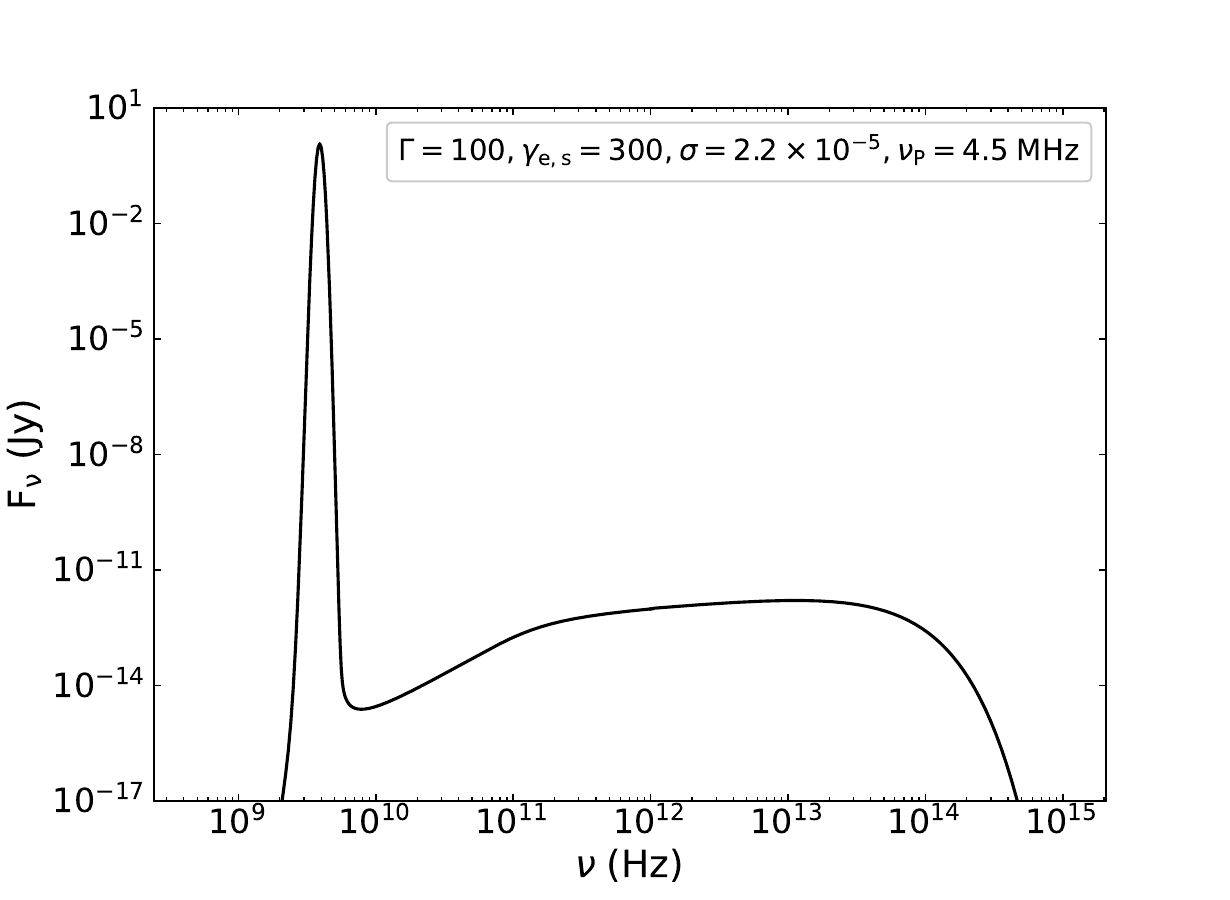}
    \includegraphics[width=\columnwidth]{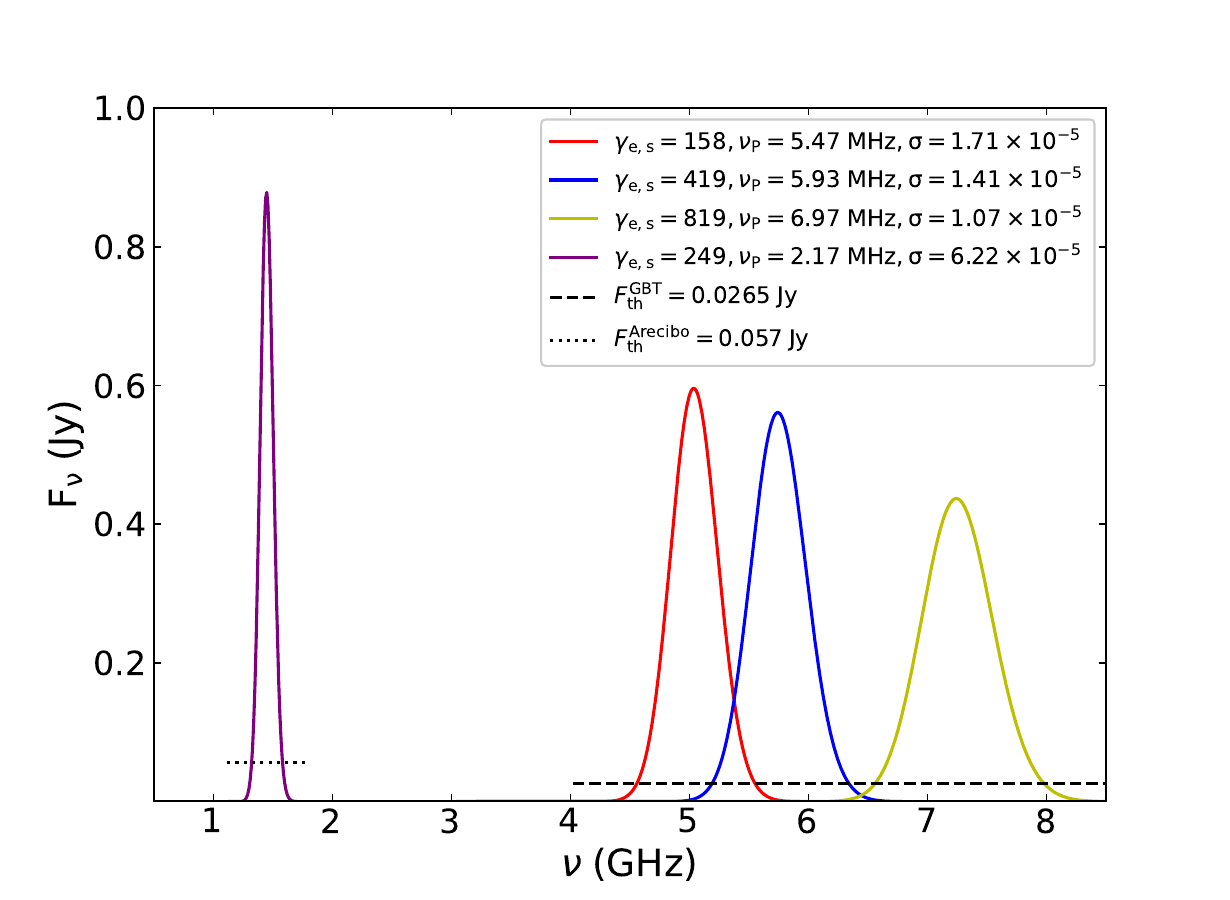}
    \centering
    \caption{Flux density of synchrotron radiation as a function of frequency for FRBs, assuming a monoenergetic electron population in an individual blob (top panel). The predicted ``observable'' burst spectral profiles under different parameters range from 1 to 8 GHz (bottom panel).}\label{MyFig2}
\end{figure}

The radiation intensity is given by (e.g., \citealp{2002ApJ...574..861S}) 
\begin{equation}\label{sec2 equ5}
  {I_{\nu}} = {j_{\nu}}{\Delta  }\frac{{1 - {e^{ - {\tau _{\nu }}}}}}{{{\tau _{\nu }}}}\;,
\end{equation}
where $j_{\nu}$ is the specific emissivity (see Appendix Eq. \eqref{appendix 3} for details), ${\tau _\nu } = {\alpha _\nu }\Delta $ is the optical depth and $\Delta$ is the width of the radiating region along the line of sight. The specific emissivity of a single electron plasma blob is given as
\begin{equation}\label{sec2 equ 5.5}
    j_\nu ^{\left[  \bot  \right]} = {j_0}{G^{[ \bot ]}}\left[ {\gamma _{{\rm{e}},{\rm{s}}}^2{\sigma ^{1/2}},\frac{\nu }{{\nu _{\rm{R}}^*}}} \right]\;,\quad {j_0} = \frac{{\pi {m_{\rm e}}}}{{2\sqrt 3 c}}\nu _{ {\rm P}}^3\sin \chi \;.
\end{equation}
The details of the function ${G^{[ \bot ]}}$ are given in Eq. \eqref{appendix 6} in the Appendix. Therefore, the radiation flux density of an individual blob in the observer's frame can be estimated as \citep{2018pgrb.book.....Z}
\begin{equation}\label{sec2 equ6}
{F_\nu }\left( {{\nu _{{\rm{obs}}}}} \right) = \frac{{(1 + z){\Gamma ^3}j{{_{{\nu ^\prime }}^\prime }^{\left[  \bot  \right]}}\left( {{\nu ^\prime }} \right)\frac{{1 - {e^{ - \tau {{_{{\nu ^\prime }}^\prime }^{\left[  \bot  \right]}}\left( {{\nu ^\prime }} \right)}}}}{{\tau {{_{{\nu ^\prime }}^\prime }^{\left[  \bot  \right]}}\left( {{\nu ^\prime }} \right)}}{V^\prime }}}{{D_{\rm{L}}^2}}\;,
\end{equation} 
where the prime means that the corresponding quantities are measured in the comoving frame, $z$ is the redshift, $V^\prime$ is the volume, and $D_{\rm L}$ is the luminosity distance. The peak frequency of emission in the observer's frame can be estimated as  
\begin{equation}\label{sec2 equ7}
    {\nu _{\rm{pk}}} = (1 + z)^{-1}0.70{\sigma ^{ - 1/4}}{\nu _{\rm{P}}\Gamma } = 0.70{(1 + z)^{ - 1}}{\Gamma _2}\sigma _{ - 4}^{ - 1/4}{\nu _{{\rm{P}},{\rm{6}}}}\;{\rm{GHz}}\;,
\end{equation}  where notation $Q_n=Q/10^{n}$ is adopted in the cgs units. In our model, the volume and optical depth of a single blob are denoted as $4\pi {R^\prime}^3/3$ and $\alpha _\nu ^{\left[  \bot  \right]}R'$, respectively. 
Taking  $\gamma_{\rm e,s}=300$, $\sigma=2.2\times10^{-5}$, $\Gamma = 100$, ${\nu_{\rm P}} = 4.5\;{\rm MHz}$, $D_{\rm L}=1$ Gpc ($z=0.193$), and $t=1\; {\rm ms}$, we derive the radiating spectrum as shown in the top panel of Fig. \ref{MyFig2}. Its peak frequency is ${\nu} _{\rm pk}=3.85\;{\rm {GHz}}$ and its peak flux density is ${F_{{\nu _{\rm pk}}}} = 1.13$ Jy. 
One can observe that the synchrotron maser emission can generate extremely narrow spectra and exceptionally bright signals (with peak flux density exceeding the emission in other bands by more than 12 orders of magnitude) at GHz frequency. For illustrating the spectral shapes in the energy bands of the GBT and Arecibo telescopes, Fig. \ref{MyFig2} (bottom) also shows the predicted spectra above the detection thresholds of the GBT ($F_{\rm limit}=0.0265$ Jy) and the Arecibo telescope ($F_{\rm limit}=0.057$ Jy) by employing three parameter sets as marked in the panel with fixing $\Gamma=100$. It can be observed that our model predicts the narrower spectrum with the lower peak frequency.

\section{Application to FRB 20121102A} \label{sec: simulation} 
The rich observational data across multiple frequencies of FRB 20121102A makes it a good candidate for exploring the radiation mechanism of FRBs. Its high brightness temperature (${\rm T_{\rm B}}\geq {10^{35}}{\rm K})$ indicates that its radiation mechanism must be coherent. Its redshift is $z=0.193$ (${ {D_ {\rm L}}}=972 \;{\rm {Mpc}} $) \citep{2017ApJ...834L...7T} and the typical burst duration is ${t_{\rm obs}} = 1\;{\rm ms}$. The distribution of burst energy ranges from $4\times{10^{36}}$ to ${10^{40}}\,{\rm {erg}}$ \citep{2021Natur.598..267L}. The observed spectral profile follows a Gaussian function \citep{2017ApJ...850...76L,2021ApJ...922..115A}. Interestingly, the distribution of peak frequency of FRB 20121102A observed with GBT at the C band exhibits a putative fringe pattern, and such a feature also seems to be seen in the observation with the Arecibo telescope at the L band ($1.15-1.73$ GHz) \citep{2022ApJ...941..127L}. We analyze these spectral properties with our model through Monte Carlo simulations.\par

Our simulation analysis is based on the observations of the peak frequency $\nu_{\rm pk}^{\rm obs}$ and the isotropic energy $E_{\rm iso}^{\rm obs}$ from the GBT and Arecibo telescopes. Assuming the emitting region is a pre-accelerated plasma that moves toward observers with a bulk Lorentz factor $\Gamma=100$, we explore the model parameter set $\{\gamma _{\rm e,s },\sigma,{\nu_{\rm P}}\}$ that can represent the $\nu_{\rm pk}^{\rm obs}$ and $E_{\rm iso}^{\rm obs}$ distributions observed with the GBT and Arecibo telescopes. Below, we outline our simulation procedure for the GBT observation.
\begin{enumerate}
\item We generate a $\nu _{{\rm{pk}}}^{\rm{sim}}$ value from the $\nu _{{\rm{pk}}}^{\rm{ obs}}$ distribution observed by the GBT \citep{2018ApJ...863....2G,2018ApJ...866..149Z}. \citet{2022ApJ...941..127L} fitted the $\nu^{\rm obs}_{\rm pk}$ distribution with a multi-Gaussian function. We adopt the fitting results from \citet{2022ApJ...941..127L} and generate $\nu _{{\rm{pk}}}^{\rm{sim}}$ based on the derived probability distribution. 

\item We generate a set of $\{ \gamma_{\rm {e,s}}, \sigma\}$ assuming that they are uniformly distributed in the range of $\gamma_{\rm {e,s}}\in[2,{10^3}]$ 
and $\sigma\in [\gamma _{{\rm{e,s}}  }^{ - 4} ,1)$, where the distribution range of $\sigma$ is derived from the weak magnetization condition ($\gamma _{\rm {e,s}}^2 > {\sigma^{-1/2}} > 1$). When $\gamma _{{\rm{e,s}}  }^2{\sigma ^{1/2}}>50$, we calculate the $\nu_{\rm P}$ value with Eq. \eqref{sec2 equ7}. We calculate the simulated peak flux density ($F_{\rm \nu_{pk}}^{\rm sim}$) at $\nu_{{\rm{pk}}}^{{\rm{sim}}}$ with Eq. \eqref{sec2 equ6} for the parameter set of $\{{{\gamma _{{\rm{e,s}}  }},\sigma ,{\nu_{\rm P}} }\}$. We check whether the $F_{\rm \nu_{pk}}^{\rm sim}$ value is in the range of $0.015 \,{\rm Jy} < F_{\rm \nu_{pk}}^{\rm sim} < 0.8\, {\rm Jy}$ as observed with GBT. If it does, we choose the parameter set and proceed to the next step. Otherwise, we discard this parameter set and repeat this step. 

\item We calculate the burst isotropic energy with 
\[E_{{\rm{iso}}}^{{\rm{sim}}} = \left( {{{10}^{36}}{\rm{erg}}} \right)\frac{{4\pi }}{{1 + z}}{\left( {\frac{{{D_{\rm L}}}}{{{{10}^{28}}{\rm{cm}}}}} \right)^2}\frac{{F_{\rm{\nu_{pk}}}^{{\rm{sim}}}}}{{{\rm{Jy}}}}\frac{{\nu _{\rm{pk}}^{{\rm{sim}}}}}{{{\rm{GHz}}}}\frac{{\Delta t}}{{{\rm{ms}}}}\;,\]
where $\Delta t$ is the burst duration. Our analysis takes the typical value with $\Delta t=1$ ms.

\item We repeat the above steps to generate a sample of $5\times10^4$ bursts, then extract a sub-sample of $8\times 10^3$ bursts by utilizing the accumulated probability distribution function [$\psi(E_{\rm {iso}}^{\rm {obs}})$] for $E_{\rm {iso}}^{\rm {obs}}$. To do so, we generate $8\times 10^3$ random numbers in the range of (0,1) and set them as the values of $\psi$, then identify the corresponding $E_{\rm {iso}}^{\rm {sim}}$ values with the inverse function of $\psi(E_{\rm {iso}}^{\rm {obs}})$.

\end{enumerate}\par

\begin{figure*}
\centering
\includegraphics[width=0.45\linewidth]{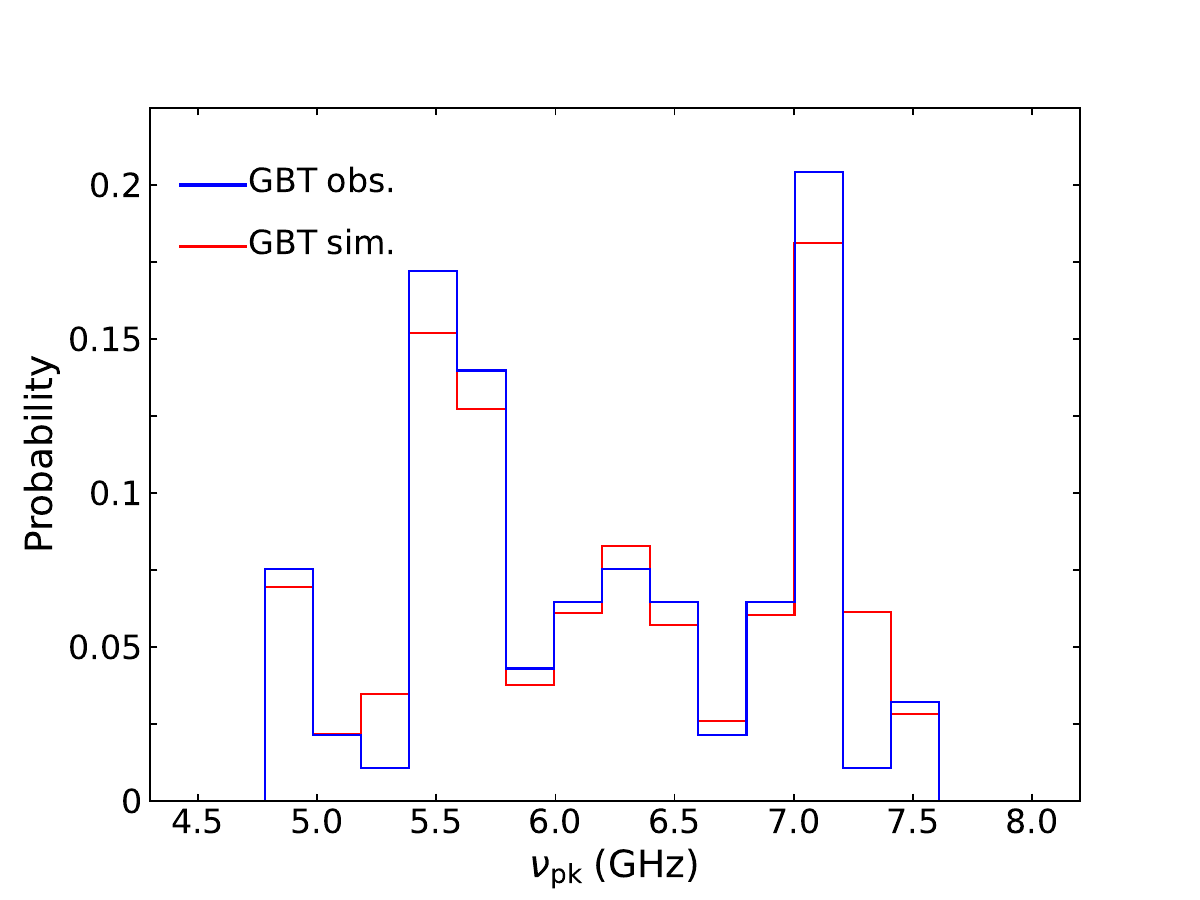}
\includegraphics[width=0.45\linewidth]{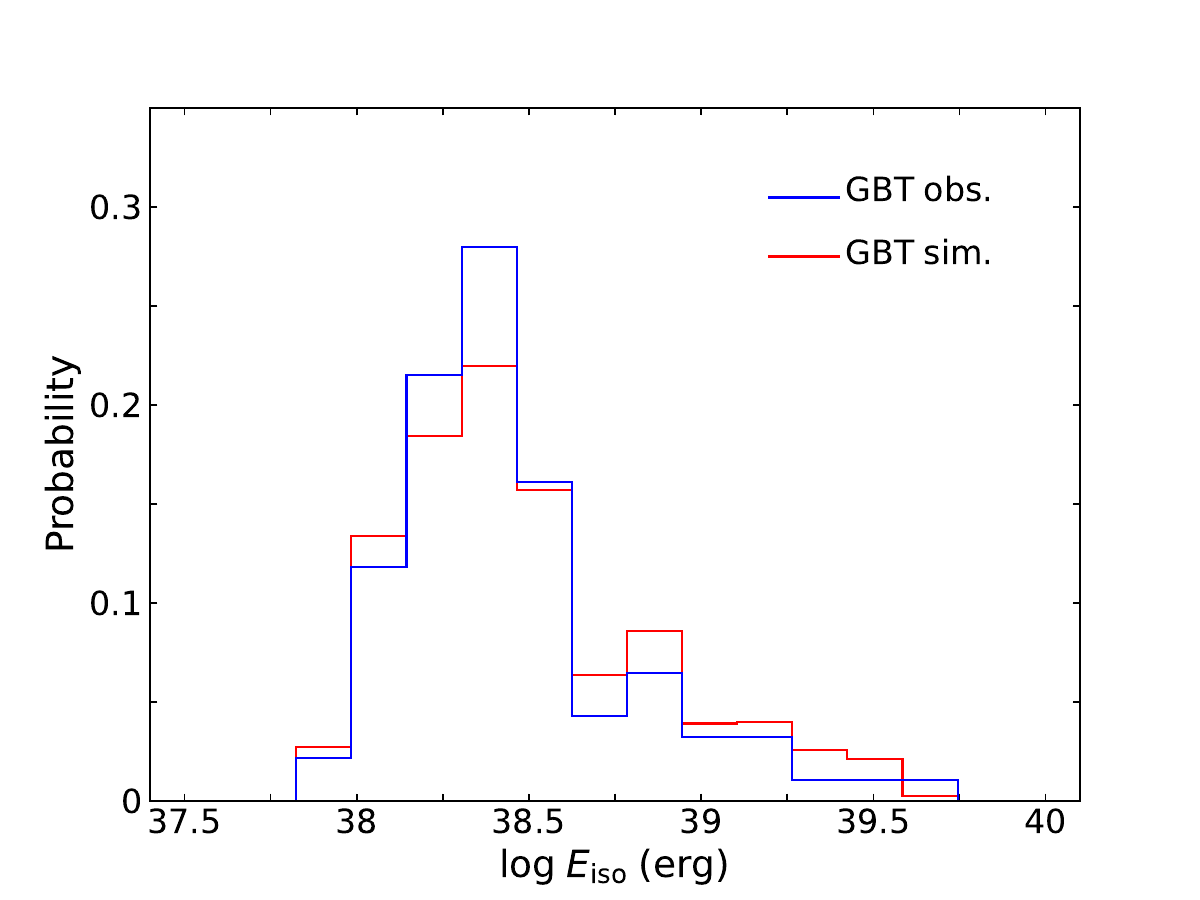}
\includegraphics[width=0.45\linewidth]{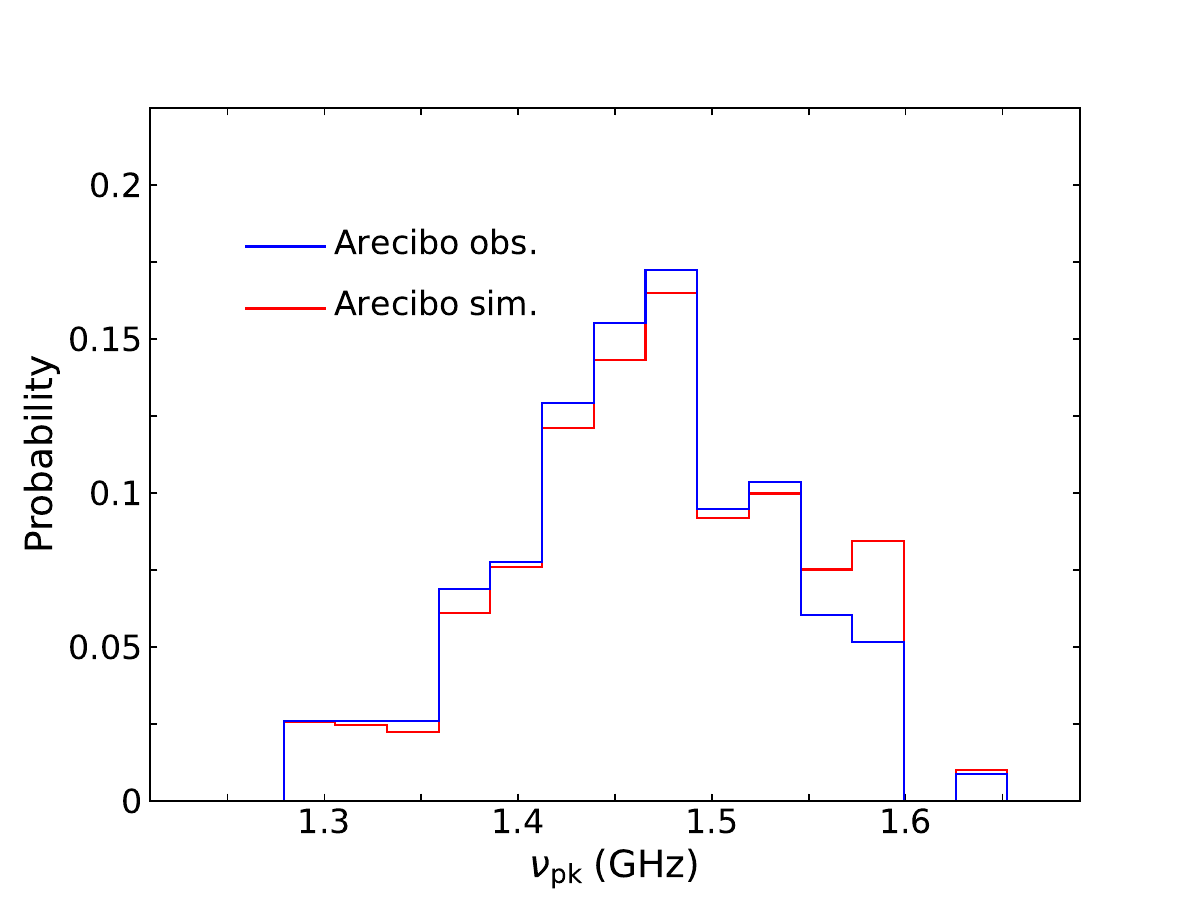}
\includegraphics[width=0.45\linewidth]{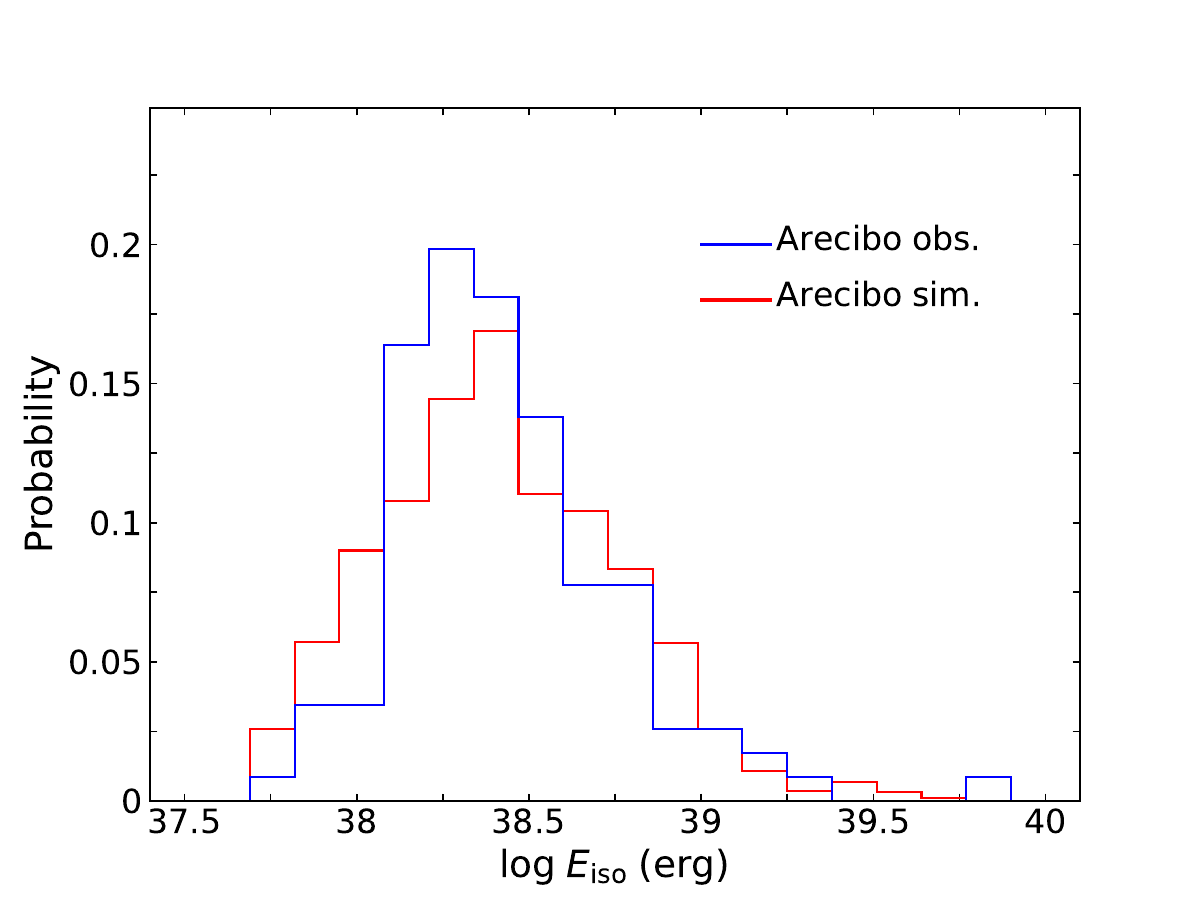}
\caption{
Comparison of the peak frequency $\nu_{\rm pk}$ and isotropic energy $E_{\rm iso}$ distributions between the simulated and observed samples, where the data of the observed samples are taken from 
\cite{2022ApJ...941..127L}.}\label{MyFig3}
\end{figure*}

Each simulated burst is characterized by $\{\nu^{\rm sim}_{\rm pk}, \, E_{\rm iso}^{\rm sim}|\gamma _{\rm e,s }, \,\sigma,  \,{\nu_{\rm P}}\}$. The parameters of the plasma, namely $\gamma _{\rm e,s}$, $\sigma$, and ${\nu_{\rm P}}$, are constrained by the consistency of the $\nu_{\rm pk}$ and $E_{\rm iso}$ distributions between the observed and simulated samples.   
Similarly, we also simulate the Arecibo telescope observations \citep{2022MNRAS.515.3577H} with the same approach. We select only those bursts whose high ($\nu_{\rm high}$) and low 
($\nu_{\rm low}$) frequencies are available because the main emission of these bursts is detected within the bandwidth of the Arecibo telescope. Since the peak flux density of these bursts is not presented in \citet{2022MNRAS.515.3577H}, we use the observed fluence in a 1 ms peak time as a proxy for the peak flux density. Figure \ref{MyFig3} compares the distributions of $\nu_{\rm pk}$ and $E_{\rm iso}$ between the observed and simulated samples. We measure their consistency with the Kolmogorov-Smirnov test. The derived p-values are larger than 0.1, indicating that the distributions of $\nu_{\rm pk}$ and $E_{\rm iso}$ between the simulated and observed samples are statistically consistent. 

\begin{figure*}
\centering
\includegraphics[width=0.33\linewidth]{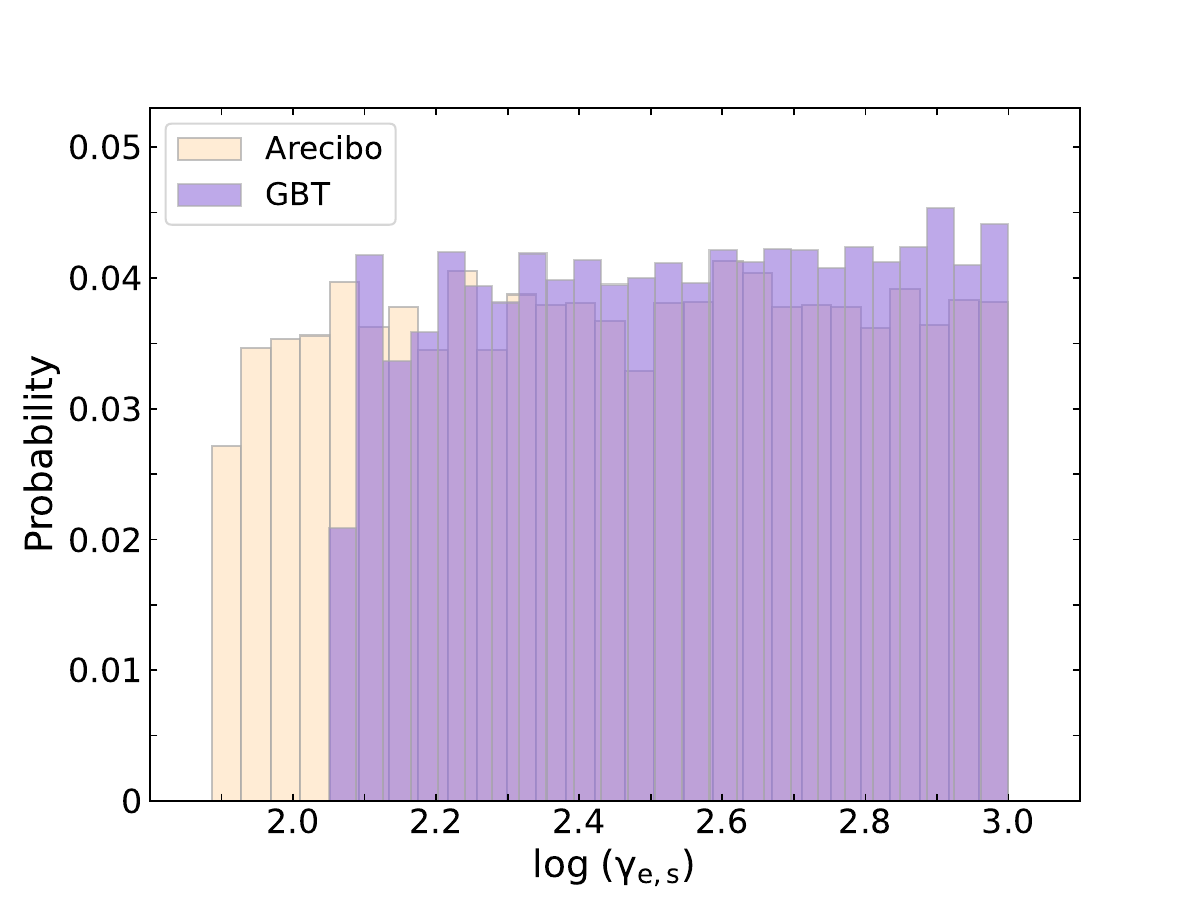}
\includegraphics[width=0.33\linewidth]{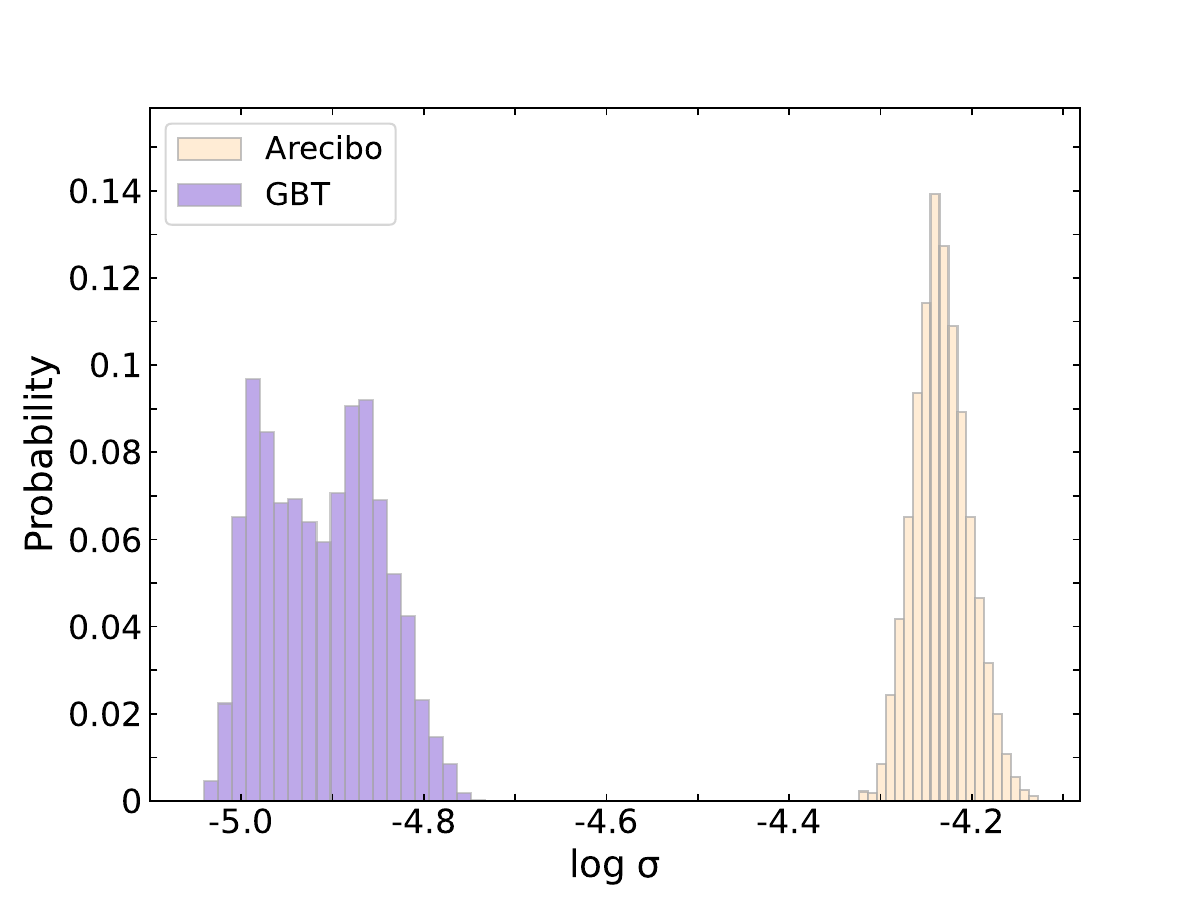}
\includegraphics[width=0.33\linewidth]{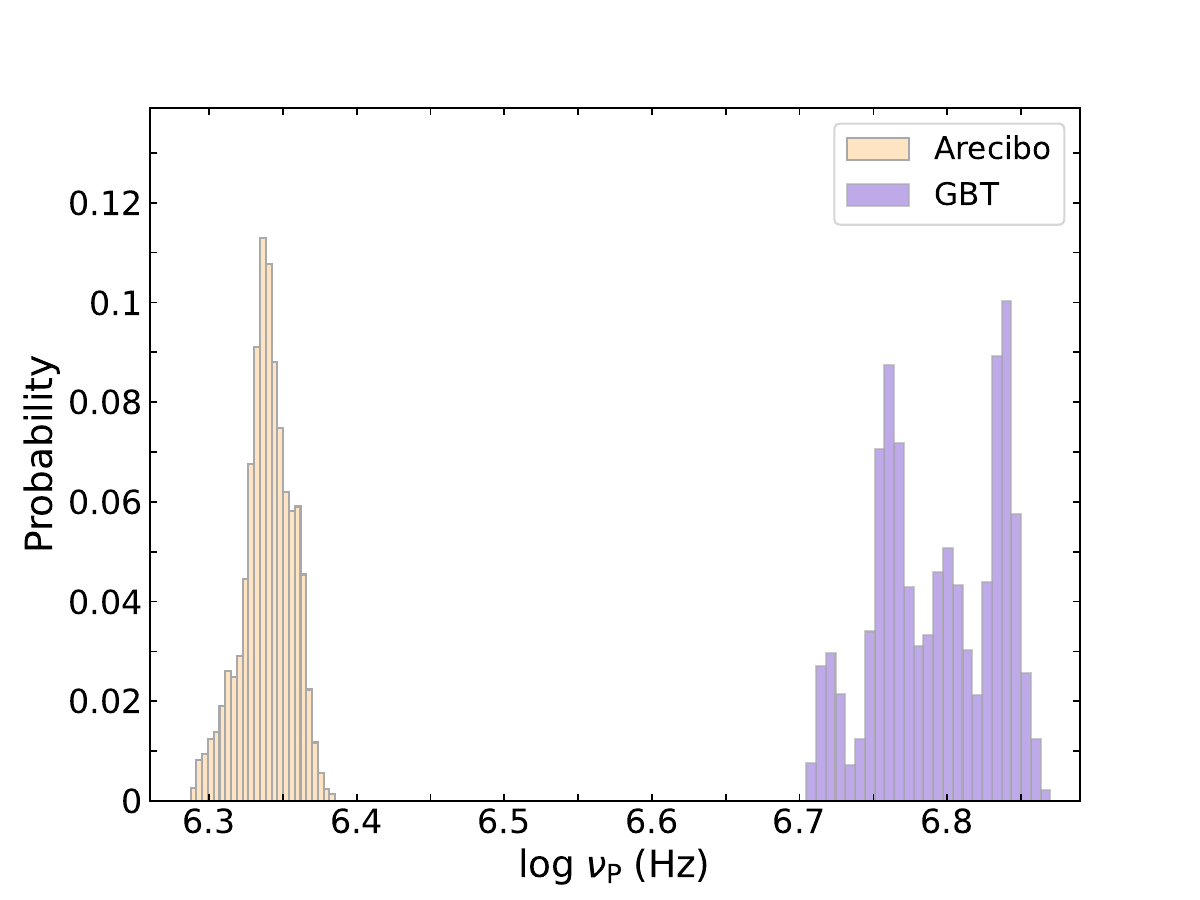}
\caption{Histograms of the model parameters $\gamma _{\rm{e,s}}$, $\sigma$ and $\nu_{\rm P}$ derived from our simulation analysis.}
\label{MyFig4}

\end{figure*}

Figure \ref{MyFig4} illustrates the histograms of the model parameters derived from our analysis. For the bursts in the $4-8$ GHz band, the $\gamma _{{\rm{e,s}}  }$ is uniformly distributed in the range of $110<\gamma _{\rm{e,s}}<3000$, and the distribution of $\sigma$ ranges from $9\times10^{-6}$ to $2\times10^{-5}$. For the bursts in the $1.15-1.73$ GHz band, the $\gamma _{{\rm{e,s}}  }$ is uniformly distributed in the range of $80<\gamma _{{\rm{e,s}}  }<3000$, and the distribution of $\sigma$ ranges from $4.7\times10^{-5}$ to $7.4\times10^{-5}$.  
The distribution of $\nu_{\rm P}$ shows several peaks in the range from $\log \nu_{\rm P}/(\rm MHz)=6.70$ ($\nu_{\rm P}=5.00 \,{\rm MHz}$) to $\log \nu_{\rm P}/(\rm MHz)=6.87$ ($\nu_{\rm P}=7.5\, {\rm MHz}$) for the bursts observed with the GBT, and a narrow peak around $\log \nu_{\rm P}/(\rm MHz)=6.35$ (2.2 MHz) for the bursts observed with the Arecibo telescope. The $\log \nu_{\rm P}$ distribution is similar to the $\nu^{\rm obs}_{\rm pk}$ distribution, suggesting that the $\nu^{\rm obs}_{\rm pk}$ of a burst is sensitive to $\nu_{\rm P}$, hence to the relativistic electron number density in the radiating region.

\section{Conclusions and discussion}\label{sec:DIS and SUM}
This paper proposes that repeating FRBs arise from synchrotron maser radiation produced by a series of local electron plasma blobs in a weakly magnetized relativistic plasma, which are induced by plasma instabilities triggered by the injected ejecta from the central engine. We present the numerical calculation of the synchrotron maser radiation spectrum in FRBs, assuming a monoenergetic electron population within an individual blob. The negative reabsorption is toward a maximum at frequency ${\nu _{\max }}=0.70\Gamma {\sigma ^{ - 1/4}}{\nu_{\rm P}}$ if $\gamma _{\rm{e,s}}^2{\sigma ^{1/2}} > 50$. The peak flux density of synchrotron maser emission is $F_{\nu_{\rm_{pk}}}\sim$ Jy and the corresponding $\nu_{\rm pk}$ is at several GHz, if $\sigma\sim 10^{-5}$, $\gamma_{\rm e,s}>100$, $\nu_{\rm P}>2$ MHz, and $\Gamma=100$. Our model predicts that FRBs with lower peak frequencies have narrower intrinsic radiation spectra. We utilize our model to account for the observed $\nu_{\rm pk}$ and $ E_{\rm iso }$ characteristics of FRB 20121102A observed with the GBT and the Arecibo telescopes. Our analysis reveals that the $\nu_{\rm P}$ distribution exhibits several peaks, which is similar to the $\nu^{\rm obs}_{\rm pk}$ distributions. 
This implies that the $\nu^{\rm obs}_{\rm pk}$ of a burst is sensitive to $\nu_{\rm P}$, which represents the relativistic electron number density. 
\par

The synchrotron maser radiations result from the negative synchrotron self-absorption if the refractive index of the relativistic plasma is less than unity. It should be noted that the refractive index in a relativistic plasma depends on the energy and angular distribution of particles \citep{1984oep.....9.2444A}. In a weakly magnetized relativistic plasma environment, the refractive index of $\rm n^2 = 1 - (\frac{\nu_P}{\nu})^2$ is valid for monoenergetic distribution and power law distribution of the electrons \citep{2002ApJ...574..861S}. Additionally, we can use the numerical results of the reabsorption coefficient to test the self-consistency of our calculations. As shown in \citet{1989aetp.book.....G} and \citet{2002ApJ...574..861S}, synchrotron maser emission is linearly polarized, if the conditions of $\left| {1 - {\rm n}} \right| \gg \left|c{\alpha _\nu }/\nu \right| $ and $\left| {{{\Delta \rm {n}}}} \right| \ll \left| {\alpha _\nu }c/\nu  \right|$ are satisfied, where $\left| {{{\Delta \rm {n}}}} \right|$ represents the difference in the refractive index between the circularly polarized modes introduced by the magnetic field. The synchrotron maser emission sharply peaks at $\nu\sim 0.70\nu _{\rm R}^*$. Thus, we have $\left| {1 - {\rm n}} \right| \sim \frac{{\nu_{\rm P}^2}}{{2{\nu ^2}}} = 1.02{\sigma ^{  1/2}}$, $\Delta {\rm n} \sim \nu_{\rm P}^2{\nu _B}/{\nu ^3} = 2.91{\sigma ^{5/4}}$, and ${\alpha _\nu }c/\nu  = 0.37\sigma $. Our analysis yields $\sigma\sim 10^{-5}$, indicating that these conditions are met. Thus, the maser emission is linearly polarized consistent with the observation of FRB 20121102A \citep{2018Natur.553..182M}.
\par
\begin{figure}
	\includegraphics[width=\columnwidth]{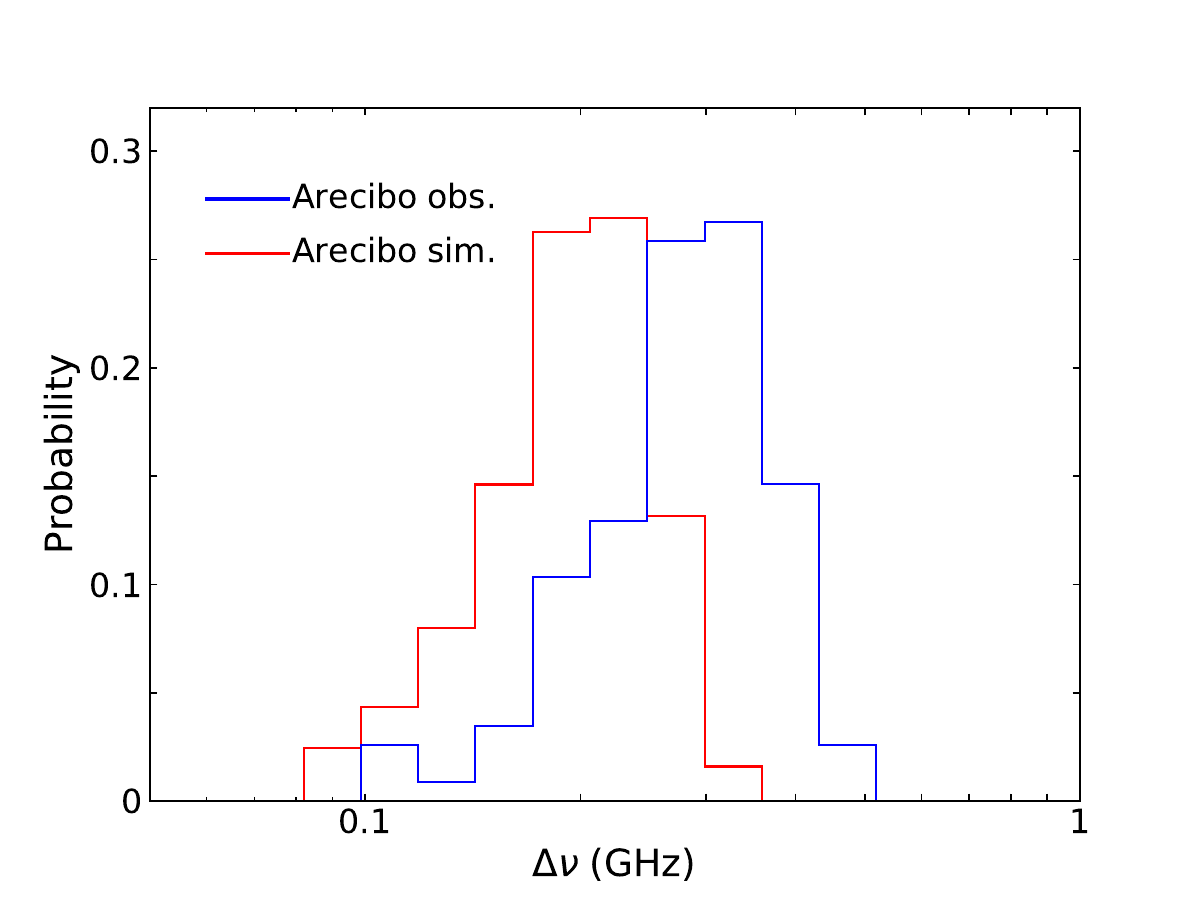}
    \centering
    \caption{Comparison of the spectral width $\Delta \nu$ of the observed sample detected by Arecibo telescope in \citet{2022MNRAS.515.3577H} with the simulated  ones.}\label{MyFig5}
\end{figure}
The $\Delta \nu$ in the sample observed with the Arecibo telescope is available. We also calculate the simulated $\Delta \nu$ for the Arecibo sample and compare its distribution with the observed ones in Fig. \ref{MyFig5}. It was found that they have similar distribution profiles, but the simulated $\Delta \nu$ distribution peaks at 0.2 GHz, while the observed peaks at 0.3 GHz. This discrepancy is possibly due to the propagation process broadening the spectra.
\par

FRB 20121102A is located in a star-forming region of a dwarf galaxy \citep{2017ApJ...834L...7T}, and in an extremely dynamic magnetized ionized environment (high Faraday rotation measure $\rm RM\sim{10^5} \;rad\;{m^{ - 2}}$, \citealt{2018Natur.553..182M}). It is spatially associated with a compact, persistent radio source (PRS) (\citealt{2017Natur.541...58C}). It also has a possible periodicity of approximately 160 days \citep{2020MNRAS.495.3551R,2021MNRAS.500..448C}.
Our analysis suggests that the synchrotron maser emission for FRB 20121102A is stimulated in a weakly magnetized region ($\sigma\sim10^{-5}$). A small $\sigma$ is primarily caused by a high relativistic electron number density. We approximate the electron population for generating the synchrotron maser emission as single-
energy electrons and the derived relativistic electron density of $n_{\rm e}={10^6} \sim {10^8}\,\rm {cm^{-3}}$. The hot corona region around accretion disk near compact objects (e.g., \citealt{2019ApJ...875..126G}), in a binary system where compact stars undergo an accretion-induced explosion or episodic jets caused by accretion (e.g., \citealt{2018ApJ...864L..12L,2021ApJ...922...98D}), or in flare of weakly magnetized material generated by magnetars \citep{2022ApJ...927....2K}, can offer such an environment. 
\par

The distribution of $\nu_{\rm pk}$ observed in FRB 20121102A displays a fringe pattern\footnote{Notably, the spectral fringe feature, observed within the range of $4-8$ GHz, was derived from a statistical analysis of 93 bursts observed in 6 hours. Despite extensive multiwavelength monitoring campaigns have been conducted on FRB 20121102A using various telescopes (e.g., \citealt{2016ApJ...833..177S,2017ApJ...850...76L,2019ApJ...877L..19G, 2019A&A...623A..42H, 2020ApJ...905L..27P}), only one burst was simultaneously detected by Arecibo at 1.4 GHz and VLA at 3 GHz \citep{2017ApJ...850...76L}.
}. Our model explains the observed fringe pattern with the inhomogeneity of relativistic electron density in plasma blobs. We note that \cite{2022Sci...375.1266F} suggested that the significant frequency-dependent depolarization at frequencies lower than 3.5 GHz in FRB 20121102A is caused by multipath propagation of the FRB emission through an inhomogenous magnetic-ionic environment. 
The $\nu_{\rm pk}$ distribution of FRB 20201102A is similar to the zebra radio spectrum typically seen in individual radio bursts emitted by the Sun and Crab \citep{2007ApJ...670..693H,2013A&A...552A..90K}. The zebra radio spectra of the solar or Crab radiations may also arise from a plasma radiation mechanism driven by uneven plasma density. \cite{2013A&A...552A..90K} proposed that the plasma density accumulation in different regions could be modulated by the magnetohydrodynamic waves in the radiation region. In addition, some energy release processes may form traps that can confine the plasma, similar to the magnetic trap proposed by \citet{2019ApJ...883...49K}. Furthermore, if low-density cavities exist within the plasma, they could impose a discrete frequency structure on the radiation \citep{2016ApJ...833...47H}. 
\par 

The extensively discussed radiation models of FRBs can generally be divided into two categories: the ``close-in'' scenario, where the emission originates from coherent curvature radiation in the magnetosphere of magnetars (e.g., \citealt{2017MNRAS.468.2726K,2018ApJ...868...31Y, 2020MNRAS.498.1397L}), and the ``far-away'' scenario, where the emission originates from synchrotron maser radiation in a relativistic outflow (e.g., \citealt{2014MNRAS.442L...9L,2017ApJ...842...34W,2019MNRAS.485.4091M,2020ApJ...896..142B,2022ApJ...927....2K}). Our model is similar to the far-away scenario. Our model suggests that the inhomogeneity of the local relativistic plasma blobs is stimulated by the injection from the central engine. 
In our model, the ejecta is required to be highly relativistic, and it is likely similar to the magnetar flares from an active magnetar \citep{2014MNRAS.442L...9L,2019MNRAS.485.4091M,2022ApJ...927....2K}.
\par

Some repeating FRB sources show complex polarization behaviors, including frequency-dependent depolarization, variation of RM, and oscillating spectral structures of polarized components \citep{2022Sci...375.1266F}. As reported by \cite{2022Sci...375.1266F}, the observed frequency-dependent depolarization and correlation between RM scatter and the temporal scattering time can be explained by multiple-path propagation through a complex environment, such as a supernova remnant-like, inhomogeneous, magnetized plasma screen (nebula) close to a repeating FRB source \citep{2022ApJ...928L..16Y}. This suggests that the temporal scattering and RM scatter originate from the same site. The model proposed by \cite{2022ApJ...928L..16Y} attributes the observed polarization variation to the propagation of the FRB emission in the nebula. They also suggested that the nebula can be responsible for the associated PRS emission.
Differently, we argue that the outbursts and propagation of FRB bursts are in the same site, which is an inhomogeneous, weakly-magnetized plasma shell. 
\par

The repeating behavior of FRBs is attributed to the episodic activity of the central engine in the framework of our model (e.g., \citealt{2019MNRAS.485.4091M}). Observations show that the duration of an outburst episode lasts from days to several months and the burst rates are erratic. For example, FRB 20201124A is a very active FRB. The follow-up observation with the FAST telescope detected a significant activity, with the detection of 1863 bursts in a period from April 1 to June 11, 2021 \citep{2022Natur.609..685X}. Four months later, another active episode was also observed by FAST with detection of over 600 bursts in a period of 4 days \citep{2022RAA....22l4001Z,2022RAA....22l4002Z,2022RAA....22l4003J,2022RAA....22l4004N}. Interestingly, comparison of the observed burst property distributions between the two burst episodes, including the spectral width, burst energy, and peak frequency, shows that they are statistically consistent. This may hint the bursts of the two episodes are from the same radiating region.
\par

\begin{acknowledgements}
We thank the anonymous referee for helpful comments. 
We thank the helpful discussions with Yuan-Pei Yang, Yao Chen, Shu-Qing Zhong, Qi Wang and Ying Gu. We acknowledge the use of the public data from the FAST/FRB Key Project. This work is supported by National Key R\&D Program (2024YFA1611700) and the National Natural Science Foundation of China (grant Nos. 12403042, 12133003, 12203013 and 12203022). E. W. L. is supported by the Guangxi Talent Program (``Highland of Innovation Talents'').
\end{acknowledgements}

\bibliographystyle{aa}
\bibliography{sample631} 

\begin{appendix}
\section{Synchrotron self-absorption and the possibility of negative reabsorption}\label{app A}
The synchrotron self-absorption coefficient obtained by the Einstein coefficient method is \citep{1958AuJPh..11..564T,1963ARA&A...1..291W}:
\begin{equation}\label{appendix 1}
 \alpha _\nu ^{\left[ p \right]} =  - \frac{1}{{4\pi {m_{\text{e}}}{\nu ^2}}}\int d {\gamma _{\text{e}}}\gamma _{\text{e}}^2P_\nu ^{\left[ p \right]}\left( {{\gamma _{\text{e}}}} \right)\frac{d}{{d{\gamma _{\text{e}}}}}\left( {\gamma _{\text{e}}^{ - 2}\frac{{d{n_{\text{e}}}}}{{d{\gamma _{\text{e}}}}}} \right)\;.
\end{equation}
Another expression is obtained according to integration by parts \citep{1967JETP...24..381Z}
\begin{equation}\label{appendix 2}
 \alpha _\nu ^{\left[ p \right]} =  \frac{1}{{4\pi {m_{\text{e}}}{\nu ^2}}}\int d {\gamma _{\text{e}}}\gamma _{\text{e}}^{ - 2}\frac{{d{n_{\text{e}}}}}{{d{\gamma _{\text{e}}}}}\frac{d}{{d{\gamma _{\text{e}}}}}\left( {\gamma _{\text{e}}^2P_\nu ^{\left[ p \right]}\left( {{\gamma _{\text{e}}}} \right)} \right) \;
\end{equation} and the specific emissivity is defined as
\begin{equation}\label{appendix 3}
j_\nu ^{\left[ p \right]} = \int {\frac{{P_\nu ^{\left[ p \right]}({\gamma _{\rm {e}}})}}{{4\pi }}\frac{{d{n_{\rm {e}}}}}{{d{\gamma _{\rm {e}}}}}} d{\gamma _{\rm {e}}}\;,
\end{equation}
where $\nu$ is the frequency, $\gamma_{\rm e}$ is the Lorentz factor of the electron, $P_\nu ^{\left[ p \right]}\left( {{\gamma _{\rm {e}}}} \right)$ is the radiation power per unit frequency emitted by a single electron with Lorentz factor $\gamma_{\rm e}$ in a polarization $\left[ p \right]$, and $\frac{{d{n_{\rm{e}}}}}{{d{\gamma _{\rm{e}}}}}$ is the isotropic distribution function of electrons. When the effect of plasma is considered,
\begin{equation}\label{appendix 4}
P_{\nu}^{[p]}\left(\gamma_{ \rm e}\right)=\frac{2 \pi {\rm e}^{2} \nu_{c}}{\sqrt{3} \gamma_{\rm e}^{2} c} S_{\nu}^{-1 / 2}\left(\gamma_{\rm e}\right)\left[x f^{[p]}(x)\right]_{x=S_{\nu}^{3 / 2} \nu / \nu_{c}}
\end{equation} 
with \[f^{[\perp, \|]}(x)= \pm K_{2 / 3}(x)+\int_{x}^{\infty} d y K_{5 / 3}(y)\; , \]
\[\nu_{c}=\gamma_{\rm e}^{3} \nu_{B} \sin \chi=\gamma_{\rm e}^{2} \frac{3 {\rm e} B \sin \chi}{4 \pi m_{\rm e} c}\;, \quad S_{\nu}\left(\gamma_{\rm e}\right)=1+\left(\frac{\gamma_{\rm e} \nu_{\rm P}}{\nu}\right)^{2} \;\]
and $\chi$ is the electron pitch angle, $K_{5/3}$ and $K_{2/3}$ is the modified Bessel function. In the case of electron distribution as a delta function, $\frac{{d{n_{\rm{e}}}}}{{d{\gamma _{\rm{e}}}}} = \delta \left( {{\gamma _{\rm e}} - {\gamma _{\rm e,s}}} \right)$, the self-absorption coefficient can be estimated as \citep{2017ApJ...842...34W}
\begin{equation}\label{appendix 5}
    \alpha _\nu ^{[p]}(g,y) = 2{\alpha _0}{y^{ - 3}}\left[ {{f^{[p]}}(x) + \left( {\frac{1}{2} - \frac{{{y^2}}}{g}} \right)x\frac{{df(x)^{[p]}}}{{dx}}} \right]\;
\end{equation}
with \[x = y{{\left( {{g^{ - 1}} + {y^{ - 2}}} \right)}^{3/2}}/\sin \chi ,\;{\alpha _0} = \frac{{\pi {\nu _B}}}{{2\sqrt 3 c}}\sqrt {\frac{{{\nu _B}}}{{{\nu_{\rm P}}}}} \sin \chi\]
and the specific emissivity can be estimated as
\begin{equation}\label{appendix 6}
j_\nu ^{\left[ p \right]}(g,y) = {j_0}{(1 + \frac{g}{{{y^2}}})^{ - 1/2}}g\left[ {x{f^{\left[ p \right]}}\left( x \right)} \right] 
\end{equation}
with \[x = y{{\left( {{g^{ - 1}} + {y^{ - 2}}} \right)}^{3/2}}/\sin \chi ,\;{j_0} = \frac{{\pi {m_e}}}{{2\sqrt 3 c}}\nu _{\rm P}^3\sin \chi \;.\]
When the effect of plasma is ignored, as in the case of a vacuum, 
\begin{equation}\label{appendix 7}
   P{{_\nu ^{[p]}}^\prime }\left( {{\gamma _{\rm e}}} \right) = \frac{{2\pi {{\rm e}^2}{\nu_{c}^\prime}}}{{\sqrt 3 \gamma _{\rm e}^2c}}{\left[ {x^\prime{f^{[p]}}(x^\prime)} \right]_{x^\prime = \nu /{\nu_{c}^\prime}}}
\end{equation} 
with \[f^{[p]}(x^\prime)= \pm K_{2 / 3}(x^\prime)+\int_{x^\prime}^{\infty} d y K_{5 / 3}(y) \;, \]
\[\nu_{c}^\prime=\gamma_{\rm e}^{3} \nu_{B} \sin \chi=\gamma_{\rm e}^{2} \frac{3 {\rm e} B \sin \chi}{4 \pi m_{\rm e} c}\;\]
and $\chi$ is the electron pitch angle, $K_{5/3}$ and $K_{2/3}$ is the modified Bessel function. The self-absorption coefficient produced by the electron distribution as a delta function can be estimated as
\begin{equation}\label{appendix 8}
\alpha {_\nu ^{[p]} }^\prime(g,y) = 2{\alpha _0}^\prime {y^{ - 3}}\left[ { - \frac{{{y^2}}}{g}x'{\frac{df(x')^{[p]}}{dx'}
}} \right] 
\end{equation}
with \[x' = y/({g^{3/2}}\sin \chi) \;,{\alpha _0}^\prime  = \frac{{\pi {\nu _B}}}{{2\sqrt 3 c}}\sqrt {\frac{{{\nu _B}}}{{{\nu_{\rm P}}}}} \sin \chi\;,\]
 where $g = \gamma _{{\text{e,s}} }^2{\frac{{{\nu _B}}}{{{\nu_{\rm P}}}}}$, $ y = \frac{\nu }{{\nu _{\rm R}^*}}$, $\nu _{\rm R}^* = \sqrt {\frac{{{\nu_{\rm P}}}}{{{\nu _B}}}} {\nu_{\rm P}}$, $\nu _B$ is the cyclotron frequency of relativistic electrons, $\nu_{\rm P}$ is the relativistic plasma frequency.
\end{appendix}
\end{document}